\documentclass[%
reprint, 
 amsmath,amssymb,
 aps,
floatfix,
]{revtex4-2}

\usepackage{graphicx}
\usepackage{dcolumn}
\usepackage{bm}
\usepackage{booktabs}

\begin{document}

\preprint{APS/123-QED}

\title{Different relative scalings between transient forces and thermal fluctuations tune regimes of chromatin organization}

\author{Anna Coletti}
 \affiliation{Department of Mathematics, University of North Carolina at Chapel Hill, Chapel Hill, NC, 27510}
\author{Benjamin L.~Walker}%
\affiliation{Department of Mathematics, University of California, Irvine, Irvine, CA, 92697}
\author{Kerry Bloom}
 \affiliation{Department of Biology, University of North Carolina at Chapel Hill, Chapel Hill, NC, 27510}
 \author{Katherine A.~Newhall}
 \email{knewhall@unc.edu}
 \affiliation{Department of Mathematics, University of North Carolina at Chapel Hill, Chapel Hill, NC, 27510}

\date{\today}

\begin{abstract}
Within the nucleus, structural maintenance of chromosome protein complexes, namely condensin and cohesin, create an architecture to facilitate the organization and proper function of the genome.
Condensin, in addition to performing loop extrusion, creates localized clusters of chromatin in the nucleolus through transient crosslinks. 
Large-scale simulations revealed three different dynamic behaviors as a function of timescale: slow crosslinking leads to no clusters, fast crosslinking produces rigid slowly changing clusters, while intermediate timescales produce flexible clusters that mediate gene interaction. 
By mathematically analyzing different relative scalings of the two sources of stochasticity, thermal fluctuations and the force induced by the transient crosslinks, we predict these three distinct regimes of cluster behavior.
Standard time-averaging that takes the fluctuations of the transient crosslink force to zero predicts the existence of rigid clusters.
Accounting for the interaction of both fluctuations from the crosslinks and thermal noise with an effective energy landscape predicts the timescale-dependent lifetimes of flexible clusters.
No clusters are predicted when the fluctuations of the transient crosslink force are taken to be large relative to thermal fluctuations.
This mathematical perturbation analysis illuminates the importance of accounting for stochasticity in local incoherent transient forces to predict emergent complex biological behavior.
\end{abstract}

\keywords{chromosome dynamics $|$ condensin $|$ gene clusters $|$ noise-induced phenomena $|$ quasipotential }

\maketitle


\section{Introduction}

The genome has an intricate and hierarchical organization that allows cells to fit an extraordinary amount of genetic material and perform important nuclear functions within a nucleus mere micrometers in size. Within the nucleus, DNA is complexed with a variety of proteins to form the fibrous material chromatin that makes up chromosomes. Contributing to chromosomal architecture are structural maintenance of chromosome (SMC) protein complexes, namely condensin and cohesin, which facilitate chromosome segregation and gene regulation \cite{paul_condensin_2019}. Such processes aid in the compaction of genetic material within the nucleus as well as allow the cell to locate and access specific genes depending on cell cycle, cell type, or environmental cue  \cite{schalbetter_smc_2017}. In particular, condensin helps regulate the retrieval of genetic information by acting on chromatin to form stochastic gene-gene crosslinks \cite{alipour_self-organization_2012} and generate loops to induce gene mixing \cite{ganji_real-time_2018, he_statistical_2020, terakawa_condensin_2017, goloborodko_compaction_2016}.

The emergent behavior induced by condensin crosslinks has been studied computationally with polymer models of chromatin \cite{fudenberg_higher-order_2012, vasquez_polymer_2014, vasquez_entropy_2016, hult_enrichment_2017, walker_transient_2019, walker_numerical_2022} in order to more directly observe the dynamics of DNA over experimental techniques. 
By representing 5kbp of DNA as beads, linked through worm-like-chain springs to form a long polymer chromatin chain, the timescale of the random SMC crosslink binding between different beads was shown to control the formation of clusters and gene interactions \cite{hult_enrichment_2017, walker_transient_2019}. These clusters, or gene neighborhoods, reveal an underlying organizational framework that optimizes gene interactions, retains a high level of genomic compaction, and yet is highly flexible.  A similar detailed level of information is not yet available experimentally; high throughput population methods such as Chromosome Conformation Capture (Hi-C) methods lack temporal evolution information, while microscopic imaging lacks genome-level resolution \cite{van_berkum_hi-c_2010}. The ability to access both the spatial arrangement {\em and the temporal re-arrangement} of beads, which is possible in simulation, is crucial for advancing our understanding of life at the cellular level.

Building and remodeling these high-level genetic neighborhoods has the potential to reveal the underlying mechanics for the configuration of the energy landscape within our nuclei. In this paper, we reveal the mathematical mechanism that describes how fluctuations in the model lead to the formation of bead clusters and the temporal mixing of beads between clusters. 
We vary the timescale of crosslinking forces between beads to change the size of this fluctuating force relative to thermal fluctuations. These various timescales produce three different dynamic regimes. 
At fast bead crosslinking timescales, rigid, unchanging clusters form. At slow crosslinking timescales, the beads interact in an amorphic state with no clusters. At intermediate mean crosslink lifetimes, clusters can both form and exchange beads, indicative of the required gene interaction for proper function.  Tuning the timescales of crosslinking provides a unifying mechanism of genome organization that gives insight into how a large number of configurational states can be rapidly remodeled as cells encounter biological challenges.

Configurations that can be rapidly remodeled were shown to take advantage of the interaction between the crosslink binding force fluctuations and the thermal noise fluctuations to promote faster bead exchange between clusters \cite{walker_numerical_2022}.  
We build upon this work by 
incorporating the timescale of the crosslink binding into the mathematical analysis to show how different relative scalings of the two sources of stochastic fluctuations, thermal noise and the force induced by the transient bonds, predict the three distinct regimes of cluster behavior mentioned above.
Standard time-averaging that takes the fluctuations of the transient binding force to zero (fast crosslinking timescale) predicts the existence of long-lived rigid clusters.
No clusters are predicted when the fluctuations of the transient binding force are taken to be large (slow crosslinking timescale) relative to the thermal fluctuations.  When accounting for the interaction of both fluctuations from the binding and thermal noise with an effective energy landscape we predict the flexible clusters and their timescale-dependent mixing.

This work provides the mathematical mechanism underlying the timescale-dependent effects of active agents in biological systems.  The timescale controls the production of fluctuations in the forces generated by the
active agents.  The relative size of these fluctuations to the thermal fluctuations produces different emergent temporal behavior.
This mechanism is not restricted to only the chromatin dynamics studied here but is potentially applicable to a wide range of cell-level biological processes such as homology searches for genetic recombination \cite{conover_changes_2011,feinstein_single-molecule_2011,bell_reca_2016}, defensive mucus barriers emerging from polymer chains of mucin \cite{wessler_using_2016,newby_blueprint_2017,vasquez_power_2023}, and error correction through configurational state changes \cite{bloom_tension_2010,kukreja_microtubule_2020,mcainsh_principles_2023}.
Thus to deepen our understanding of how such biological systems function, we have shown it is necessary to properly analyze the interplay of different fluctuations and the {\em timescale} on which they are generated.

\section{Results}

\subsection{Timescale of crosslink binding force drives cluster formation and dynamics}

\begin{figure*}[ht]
  \includegraphics[width = 0.95\linewidth,trim={0 2cm  0 0}]{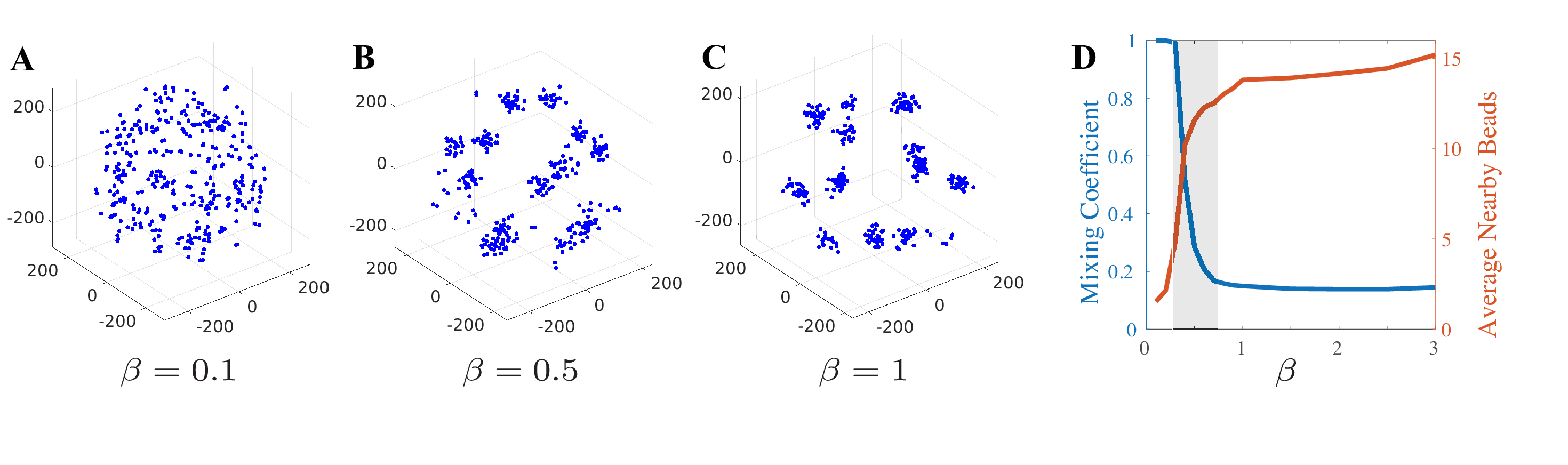}

\caption{(A-C) Snapshot of 361 bead model for increasing binding timescale. We observe three clustering regimes: (A) amorphic, (B) flexible, and (C) rigid. (D) Mixing coefficient and average nearby number of beads for a range of $\beta$-values for 361 bead model. Flexible clustering occurs for the range of $\beta$-values within the grey region. In this regime, clusters emerge and beads interact.}
\label{fig: 361mixing}
\end{figure*}

We establish that it is the timescale of the stochastically-switching SMC protein crosslinking force through random binding and unbinding that affects the organization of the beads and the dynamics of the clusters.  
To show this, we create an idealized 361-bead model that represents the nucleolus, removing the chromatin-chain springs between beads to emphasize that it is the stochastically-switching force in competition with the excluded volume force that creates the clustering structure.  The number of beads is set by the length of the biological chromatin chain with each bead representing 5 kilobase pairs of DNA. 
The beads bind and unbind based on a biologically feasible random model for the SMC protein crosslinks that is more likely to choose pairs of nearby beads to bind.

Fig.~\ref{fig: 361mixing} shows the formation of clusters when the timescale of the random crosslink binding force, $\beta$, is fast enough.
The repulsive excluded volume force and thermal fluctuations acting on the beads are in competition with the attractive stochastically changing pairwise binding forces that coalesce many beads into a cluster. 
Furthermore, changing the timescale of the crosslink binding force,  $\beta > 0$, from slow to fast results in three distinct clustering behaviors: 
The system passes from amorphic behavior (dissolution of clusters with many bead interactions) to flexible clustering (frequent bead exchanges between clusters) and then to rigid clustering (minimal bead interaction between clusters).

The observed timescale-dependent clustering behavior mirrors that seen in \cite{walker_transient_2019} which employed a larger polymer-like chromosome model and a different random model for the crosslinking proteins in the nucleolus.
This demonstrates the universality of the clustering dynamics that are driven by the timescale of the fluctuating binding forces. 


\subsection{Time-averaged force predicts existence of rigid clustering dynamics}

To mathematically determine the effective attractive force generated by the stochastic binding we work with a further reduced 3-bead model.  We start by showing that
a simple time-averaging of the stochastic binding force can predict the existence of rigid clustering dynamics but not the timescale-dependent mixing of flexible clustering dynamics. 

The chosen random model for the crosslink binding forces has the added advantage of fitting into  
 a continuous-time Markov chain (CTMC) framework.
For the idealized 3-bead model, the four Markov states are easily enumerated: all beads unbound ($s = 1$), beads 1 and 2 bound ($s = 2$), beads 1 and 3 bound ($s = 3$), and beads 2 and 3 bound ($s = 4$). The time evolution of the states follows a general CTMC process with a transition rate matrix $S$. Included in $S$ is the bead binding rate that is proportional to the bead separation distance given by affinity function $a(\cdot)$ and a constant bond breaking rate $c$.

 Now that the binding stochasticity is formulated as a Markov chain, finding the time-averaged force on each bead is straightforward \cite{pavliotis_averaging_2008}. The 
 percent of time spent in each Markov chain state
 is given by $\vec{r}(\vec x)$, the normalized null vector of the switching matrix $S(\vec x)$, so that the time-averaged force for each coordinate is
 \begin{equation}\label{eq: deterministic}
    \frac{dx_k}{dt} = \sum_{s=1}^4 v_k(\vec{x}; s) r_s(\vec{x}) = \langle v_k (\vec{x})\rangle \; \text{for} \; k= \{1, \dots , 6\}
\end{equation}
where $v_k(\vec{x}; s)$ is the force on coordinate $k$ when the system is in Markov state $s$. (Note, we have concatenated the $x$ and $y$ coordinates of each bead into a single vector $\vec{x}$.) This removes the stochasticity and results in a deterministic effective force for the system, $\langle \vec{v} \rangle$.

\begin{figure}[!ht]
  \includegraphics[width = 1\linewidth,,trim={0 0 0 0.7cm}]{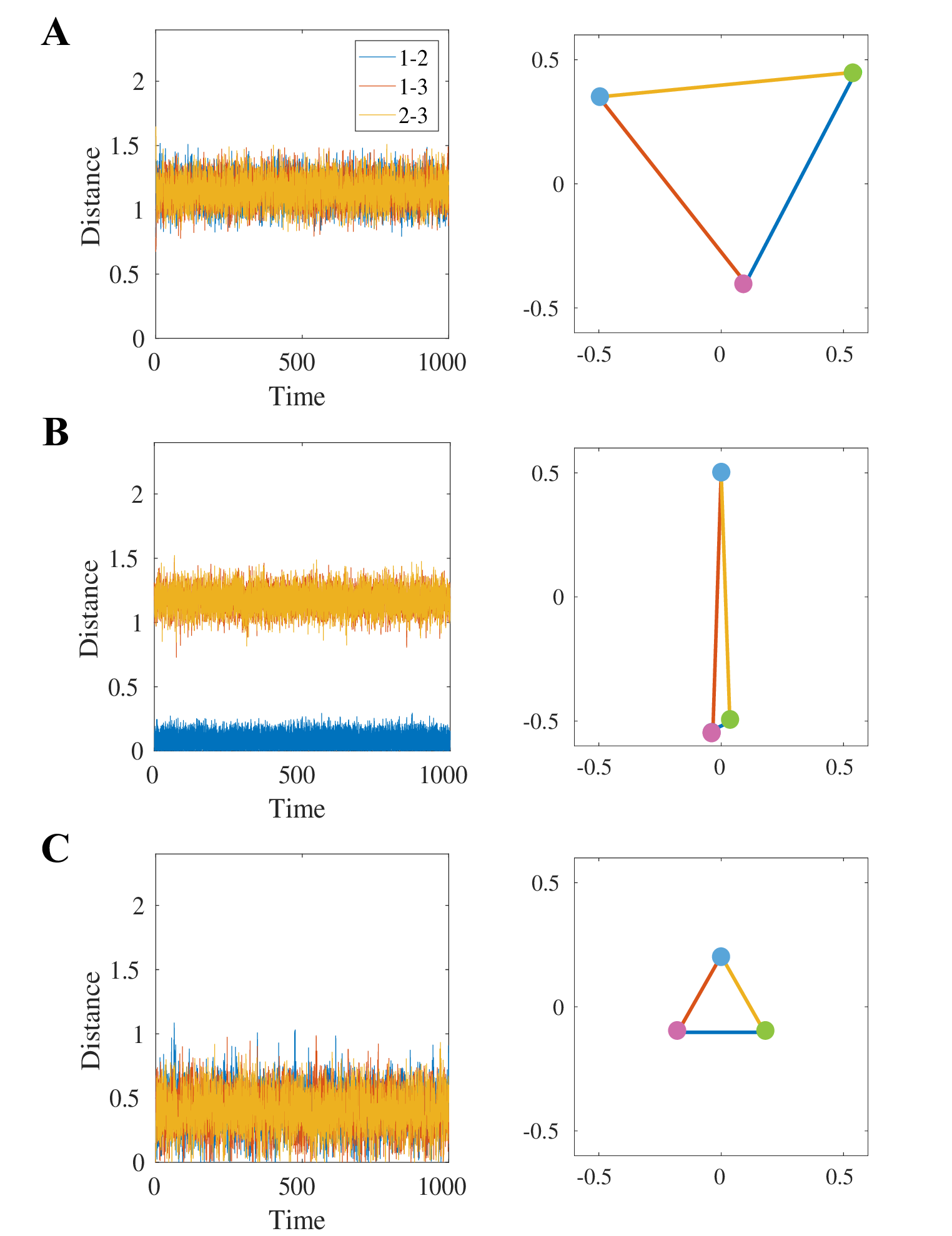}

\caption{(A-C) plots of pairwise distances of three beads over time and a snapshot of the three bead configuration: (A) all beads unbound: balance between excluded volume and confinement forces keep beads in a large triangle configuration, (B) 2-bead cluster state with beads 1-2 bound, (C) 3-bead cluster state: all beads remain close, rapid switching between pairwise bonds keeps beads in a small triangle configuration.}
\label{fig: states}
\end{figure}

The stable fixed points of \eqref{eq: deterministic} are the observed clusters. One fixed point is a 3-bead cluster in which the beads are in 
a small triangle configuration; see Fig.~\ref{fig: states}C. This is distinguished from the unbound state $s=1$, in which the beads form a triangle but with larger pairwise distances; see Fig.~\ref{fig: states}A. The other three fixed points are 2-bead clusters in which two beads are superimposed and the third is farther away and unbound; see Fig.~\ref{fig: states}B. These are distinguished from states $s=2, 3, 4$ as the clusters persist longer than any single bond lifetime. 

In addition to predicting the existence of clusters, we can test this time-averaging procedure's ability to 
predict the lifetime of a cluster. We do this by considering the effects of small perturbations of noise about the average by studying the equation,  
\begin{equation} \label{eq: vbar langevin}
    dx_k = \langle v _k (\vec{x}) \rangle dt + \sqrt{2 \epsilon}\; dB_k
\end{equation}
in the limit as $\epsilon \to 0$. Even though the vector of forces is the gradient of a potential function for each state $s$, i.e. $\vec{v}(\vec{x}; s) = -\nabla_x U(\vec{x}; s)$, it is not true that $\langle \vec{v}  (\vec{x})\rangle = -\nabla_x \langle U (\vec{x}) \rangle $ since the CTMC null vector $\vec{r}$ depends on the positions $\vec x$. However, we numerically find an effective potential $U_{\text{eff}}$ such that $-\nabla_x U_{\text{eff}} (\vec{x}) \approx \langle \vec{v}  (\vec{x})\rangle$ along the most probable transition path connecting a minimum of $U_{\text{eff}}$ (recall these minima are the stable fixed points of \eqref{eq: deterministic}) to a saddle point using the String Method~\cite{e_string_2002}. We asymptotically approximate the escape times from a cluster using the well-known Arrhenius law, 
\begin{equation}\label{eq: Arrhenius}
\log(\mathbb{E}[\tau]) \sim \frac{\Delta U_{\text{eff}}}{\epsilon}
\end{equation}
in the limit as $\epsilon \to 0$, where $\Delta U_{\text{eff}}$ is the change in the effective potential generated by the time-averaged force along the most probable transition path \cite{zhou_quasi-potential_2012}.

The time-averaged force, and thus its effective potential, is invariant to the overall timescale of the transition rate matrix. Scaling $S$ by some constant $\alpha$ does not affect the fraction of time spent in each state because $\vec{r}$ is also in the null space of $\alpha S$. Thus, Eqs. (\ref{eq: deterministic}) and (\ref{eq: vbar langevin}) are unchanged when scaling the transition rate matrix $S$ by $\alpha$. The escape times from a cluster predicted by \eqref{eq: Arrhenius} are valid for describing the rigid clustering dynamics that are largely invariant to the binding timescale parameter $\alpha$.
As we will later show, this regime has the highest effective energy barrier and thus the longest-lived clusters.

\subsection{Relative scalings of fluctuations predicts three different clustering regimes}

The key to understanding clustering dynamics is accounting for
the relative size of fluctuations between the stochastic binding and the small perturbations of thermal noise. To analyze the dynamics mathematically we perform an asymptotic expansion as the size of the fluctuations goes to zero. We consider taking both fluctuations to zero at the same rate and at different rates. The latter allows one noise source to dominate. Rigid clusters arise when the
binding fluctuations are small and only thermal noise drives transitions between clusters.
 Amorphic arrangements arise when the binding fluctuations are large and take place on a longer timescale than the fast thermal fluctuations. The stochastic binding alone drives transitions between arrangements.
Taking fluctuations to zero at the same rate
 allows for interaction between the noises. We find that this interaction predicts the flexible cluster dynamics that depend on the timescale of the binding. The combination of stochastic binding and thermal fluctuations drives transitions between clusters.  Together, these three asymptotic regimes explain the three regimes seen in \cite{walker_transient_2019} and Fig.~\ref{fig: 361mixing}. In the remainder of this section, we elaborate on the mathematics and thus reveal the mechanism behind these three dynamic regimes.

We include the effects of fluctuations about the mean force so that the position of each bead follows the overdamped Langevin-like equation given by 
\begin{equation} \label{eq: langevin}
    dx_k = v_k (\vec{x}; s) dt + \sqrt{2 \epsilon} dB_k.
\end{equation}
where $s$ is the state of the CTMC. How we choose to scale the transition rate matrix $S$ determines how the relative rates of the two different fluctuations go to zero. 

We start with the scaling $\frac{1}{\epsilon^2}S$ which takes the binding fluctuations to zero faster than the perturbations from thermal noise. We seek an effective potential $W$ (also known as a quasipotential) to replace $U_{\text{eff}}$ in Eq. (\ref{eq: Arrhenius}). Therefore we assume a steady-state distribution 
\begin{equation} \label{eq: wkb ansatz}
    p_s(\vec x) = r_s(\vec x)\exp \bigg( -\frac{1}{\epsilon}W(\vec x) \bigg)
\end{equation}
as the asymptotic solution for the system of steady-state Fokker-Planck equations, 
\begin{equation} \label{eq: FP }
    \frac{\partial p_s}{\partial t} = -\sum_{k=1}^6 \frac{\partial}{\partial x_k} [v_k(\vec{x};s) p_s]+ \epsilon \sum_{k=1}^6 \frac{\partial^2}{\partial x_k^2}[p_s] + \frac{\alpha}{\epsilon^2} \sum_{j = 1}^4 S_{sj}p_j.
\end{equation}
The element of the CTMC transition matrix $S_{sj}$ gives the transition rates from state $j$ into state $s$. Thus, the last term couples the process between different states. Setting \eqref{eq: FP } to zero and plugging in \eqref{eq: wkb ansatz} yields the leading order equation
\begin{equation} \label{Sr = 0}
    S \vec{r} = 0
\end{equation}
with next order equation
\begin{equation} \label{eq: next order}
    r_s \vec{v} \cdot \nabla_x W + r_s \nabla_x W \cdot \nabla_x W = 0.
\end{equation}
Thus, $\vec{r}$ is the steady-state distribution of $S$ and summing Eq. (\ref{eq: next order}) over $s$ reveals that $-\nabla_x W = \sum_s \vec{v}(\vec{x}; s) r_s(\vec{x})$ is exactly the naive time-average from above that does not depend on $\alpha$ and predicts rigid clustering dynamics.

If instead we take the scaling $\frac{1}{\epsilon}S$, both sources of noise remain in the leading order equation given by 
\begin{equation}\label{eq: ADSr = 0}
    [A(\vec x,\nabla_x W) + D(\nabla_x W) + \alpha S(\vec x)]\vec{r} = 0
\end{equation}
for advection (drift) matrix $A$, diffusion matrix $D$, and switching rate matrix $S$; see \cite{walker_numerical_2022} for details.
In this regime, both sources of noise are important for flexible clustering. We see the effects of varying the binding timescale parameter $\alpha$, which controls how quickly the simulation switches between the Markov chain states. This role mirrors that of the kinetic timescale parameter $\mu$ in the large-scale simulations of \cite{hult_enrichment_2017, walker_transient_2019} in which $\mu$ is a ``tuning knob" used to set the kinetic timescale on which the crosslinks bind and unbind.

\begin{figure}[!htb] 
\centering 
  \includegraphics[width=1\linewidth,trim={0 0 0 3cm}]{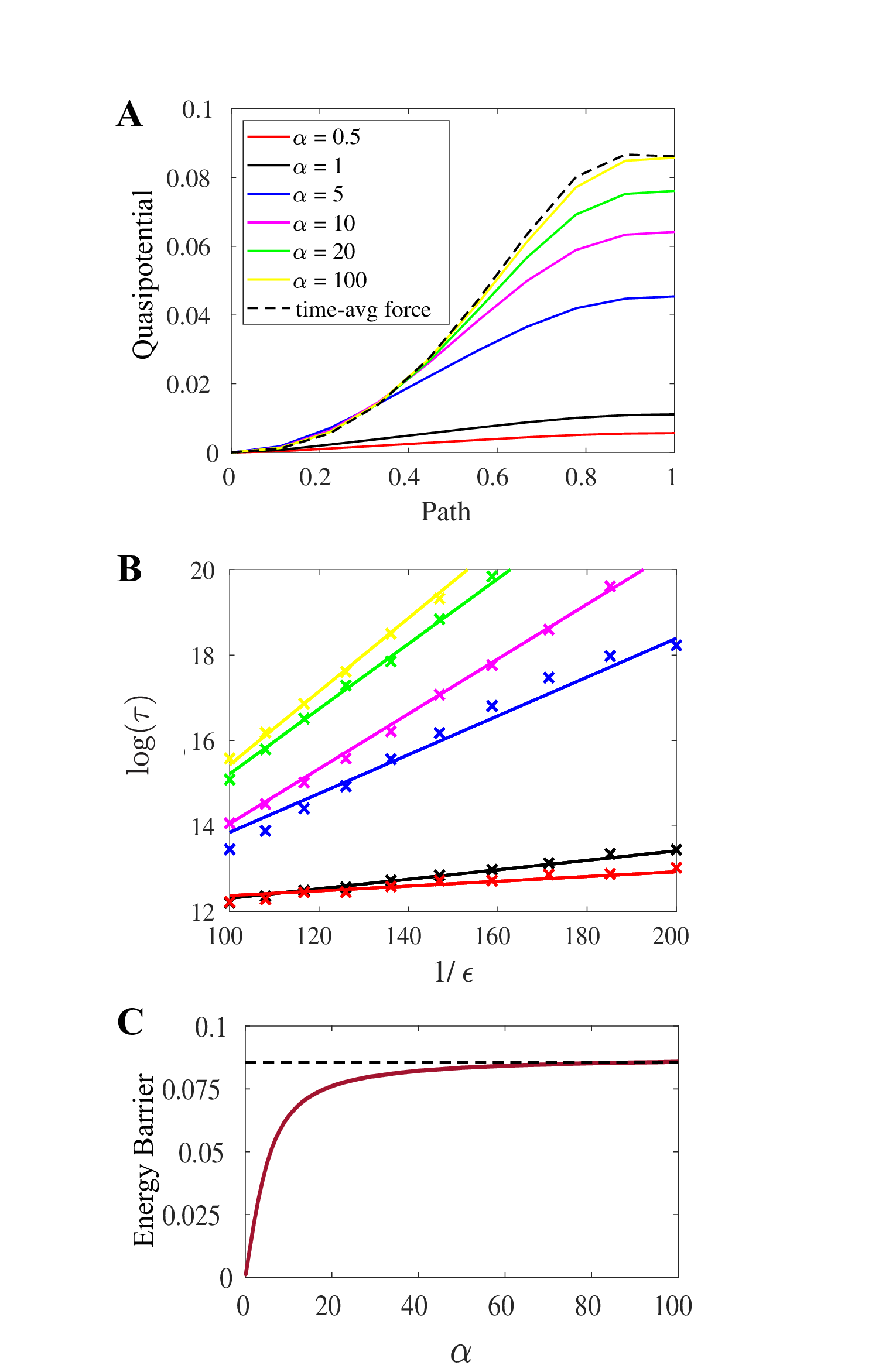}
\caption {Transition: 2-bead cluster to 2-bead cluster. (a) Quasipotential (solid) for different $\alpha$ values and for the time-averaged binding force (dotted) along the transition path, normalized to unit length. (b) Comparison of asymptotic escape times computed via Monte Carlo simulation to slope taken from quasipotential barrier height. (c) Quasipotential energy barrier height vs.~switching timescale $\alpha$. \label{fig: quasi_MC_EB} }
\end{figure}

To further illustrate the changing dynamics with $\alpha$, we compare the predicted lifetime of the cluster governed by Eq.~({\ref{eq: Arrhenius}}) with Monte-Carlo simulations. 
The predicted lifetime depends on the barrier height of the quasipotential 
 along the most probable path connecting the minimum to the saddle point.  
 Fig.~\ref{fig: quasi_MC_EB}A shows the quasipotential along such a path 
 from a 2-bead cluster to the saddle point of another 2-bead cluster for corresponding $\alpha$-values. 
The asymptotic escape times from one 2-bead cluster to another 2-bead cluster are shown in Fig.~\ref{fig: quasi_MC_EB}B. The slope value is given by the quasipotential barrier height and well expresses the linear relationship of the mean escape times. As $\alpha$ increases, the effective energy barrier approaches the barrier predicted by the naive time average discussed previously; see Fig.~\ref{fig: quasi_MC_EB}C.

\begin{figure}[!tbph]
  \centering 
  \includegraphics[width=1\linewidth,,trim={0 0 0 0.5cm}]{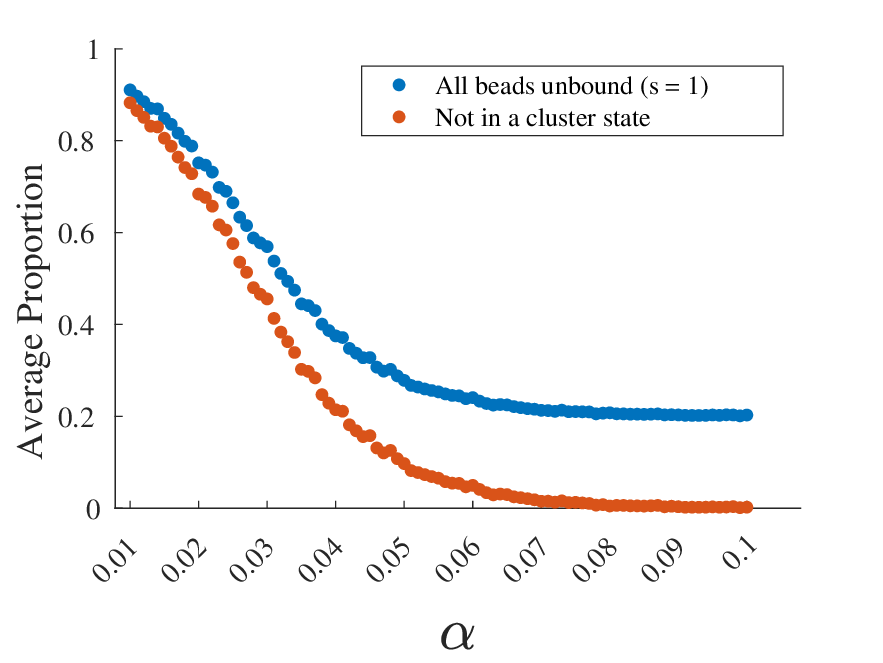}
  \caption{Average proportion of simulation spent in $s=1$ unbound state of the Markov chain and average proportion of simulation not spent in a cluster state vs. $\alpha$. \label{fig: unbound}}
\end{figure}

Finally, if we keep $S$ independent of $\epsilon$, retaining all the fluctuations, the system equilibrates within each Markov chain state and there are no clusters predicted. In this amorphic regime, the effects of the binding noise are pushed into a higher order so that the leading order does not contain any contributions from the CTMC. We observe the beads are often unbound, settling into a ``large triangle" configuration; see Fig.~\ref{fig: states}A. This long-observed state of the system is a balance between the excluded volume and the confinement forces. We do not consider this a cluster as it would dissolve in the absence of the confinement forces, leading to amorphic dynamics. Fig.~\ref{fig: unbound} shows the formation of clusters as $\alpha$ increases. The system is able to remain in a cluster state despite all the beads being unbound as the system does not have time between binding events to reach the large triangle equilibrium.

\subsection{Effects of crosslink binding fluctuations on different transitions}

\begin{figure*}[!htb]
\hspace{0.3in}
      \includegraphics[width = 1\linewidth]{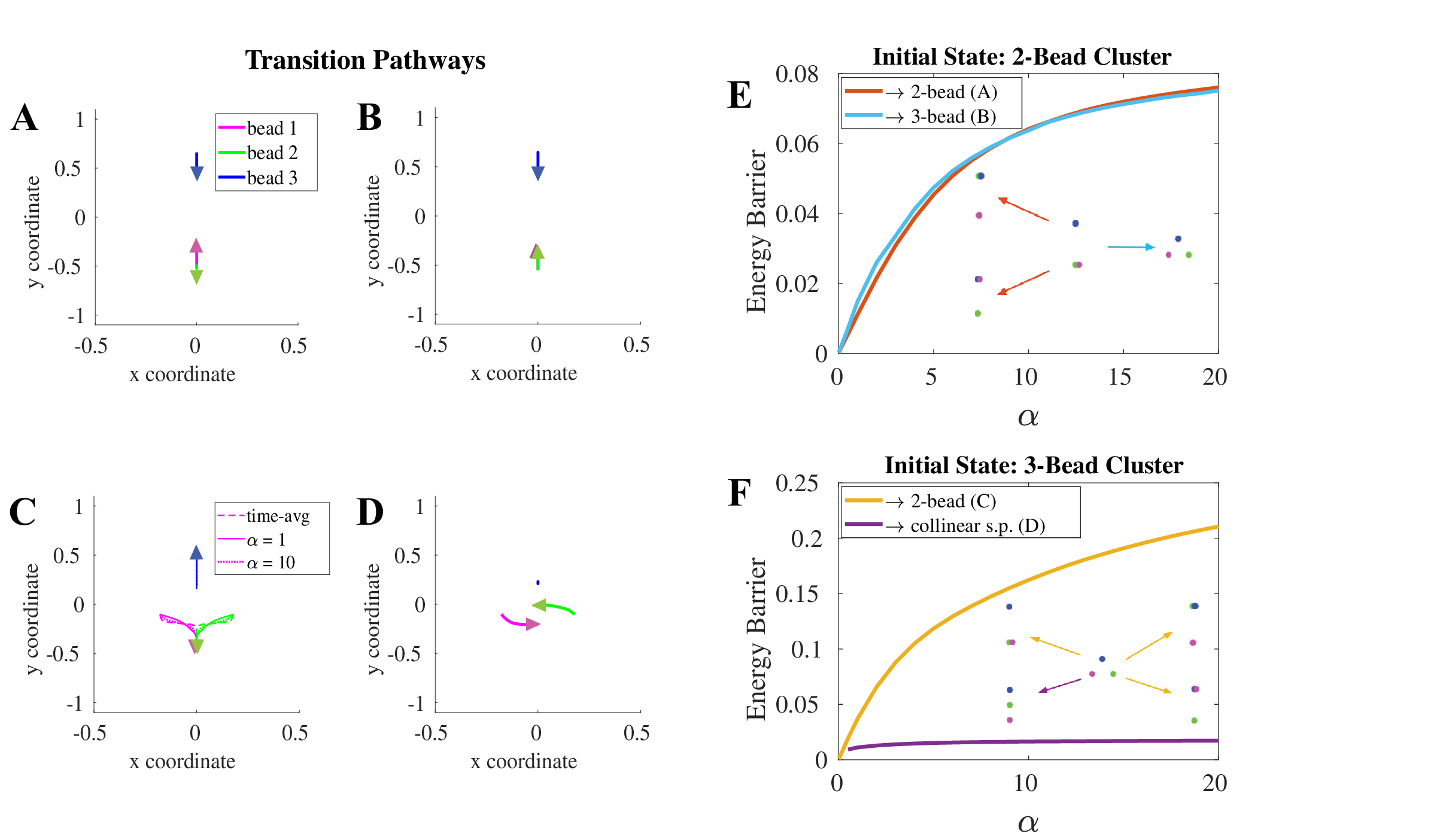}
\caption{(A-D) Transition pathways to a saddle point: (A) 2-bead cluster on the way to a different 2-bead cluster (B) 2-bead cluster on the way to a 3-bead cluster, (C) 3-bead cluster on the way to 2-bead cluster for $\alpha=1$, $\alpha=10$, and time-averaged binding force ($\alpha\to\infty$), (D) 3-bead cluster to the collinear saddle point configuration that continues to a rearrangement of the 3-bead cluster, or to a different saddle point on the way to a 2-bead cluster. (E, F) Energy Barrier heights vs. $\alpha$ for these  transition pathways}\label{fig: MPP and EB}
\end{figure*}

Scaling $S$ by $\frac{1}{\epsilon}$ revealed the 
dependence of cluster lifetimes on the
crosslink binding timescale, $\alpha$.  By accounting for both sources of randomness in the system we computed an accurate quasipotential that predicted a higher energy barrier height between minimum states as $\alpha$ increased, and thus longer cluster lifetimes.  However, this increase is not uniform across all transitions as we show in Fig.~\ref{fig: MPP and EB}.

Of the ``uphill" most probable transition paths the system takes from one minimum state to the saddle point shown in Fig.~\ref{fig: MPP and EB}A-D, only the 3-bead on its way to a 2-bead cluster has noticeable dependence on $\alpha$ (Fig.~\ref{fig: MPP and EB}C).  This transition also has the largest change in quasipotential energy barrier with $\alpha$ (Fig.~\ref{fig: MPP and EB}F) indicating removing a bead from a cluster is more sensitive to the binding timescale that rearrangements within the cluster (pathway in Fig.~\ref{fig: MPP and EB}D). The transition shown in Fig.~\ref{fig: MPP and EB}D to the ``collinear" saddle point configuration also has the lowest energy barrier height amongst the transitions, with limited dependence on $\alpha$ for either the pathway or the height.
This indicates that the binding force is not important for this transition, as  
the stochasticity of the binding does not help the system make the transition.

Starting in the 2-bead cluster state, our system can transition into a different 2-bead cluster (Fig.~\ref{fig: MPP and EB}A) or into the ``small triangle" 3-bead cluster (Fig.~\ref{fig: MPP and EB}B). These transitions have similar pathways up to each saddle point that do not significantly depend on $\alpha$ and have similar quasipotential energy barrier heights for the range of $\alpha$ values from 0 to 20. This suggests that the system will take either transition with similar probability and do so in approximately the same amount of time.

\section{Discussion}

We have shown the underlying mathematical mechanism behind how the 
relative scaling between the thermal fluctuations and
the timescale of the model condensin crosslinking force tunes the regime of chromatin organization into one of three regimes: flexible clustering, rigid clustering, and an amorphic state. 
This crosslinking timescale controls the production of random fluctuations in the force and in turn the dominating terms in the asymptotic perturbation when searching for an effective thermal equilibrium. 
This analysis emphasizes the importance of accounting for stochasticity in local incoherent transient forces to predict emergent complex
biological behavior.

Flexible clustering dynamics emerge from
accounting for fluctuations from both the stochastic binding force and the thermal noise, keeping both of their corresponding terms in the Fokker-Planck equation balanced in the limit as the size of the fluctuations goes to zero. 
To produce this interaction mathematically, we scale the binding rate matrix $S$ by $\frac{1}{\epsilon}$.
Taking $\epsilon\to 0$
with this scaling, an effective binding force remains that allows for stable clusters.  This limit also accounts for the fact that the fluctuations of the binding force aid in transitions between clusters, as it creates finite periods of time when the thermal noise has to overcome smaller forces to push the system into a new cluster state.  Since our added binding timescale parameter $\alpha$ controls the size of the fluctuations, it also controls the size of the effective energy barrier for predicting transition times between stable cluster states given by the Arrhenius law.

\begin{figure*}[t]
    \includegraphics[width = 0.90\textwidth]{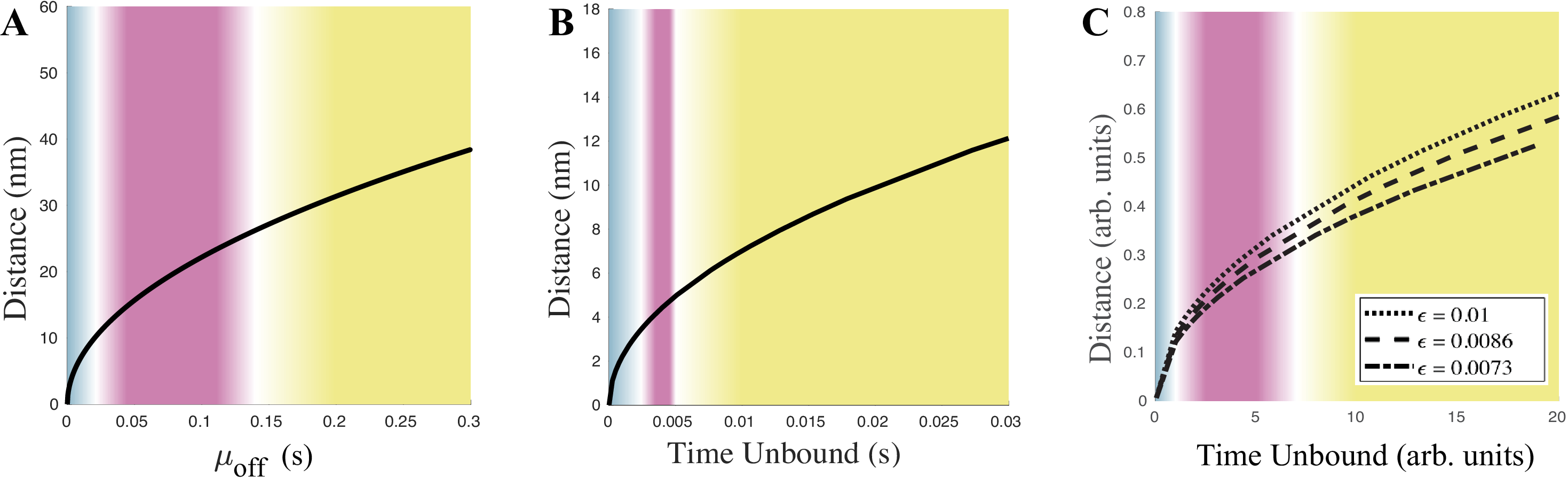}
    \caption{Typical distance a bead diffuses while unbound (black line) and general classification of the three regimes (blue rigid; pink flexible; yellow amorphic) for three different models. (A)Biologically-realistic model \cite{hult_enrichment_2017}  
    : the typical distance $\sqrt{3\frac{k_b T}{\gamma}\mu_{\text{off}}}$ beads diffuse while the crosslink is inactive; $\mu_{\text{off}}$ is the prescribed mean time for the crosslink to remain inactive. (B) 361-bead model: the typical distance $\sqrt{3\epsilon T_{\textrm{unbound}}}$ beads diffuse while unbound; $T_{\textrm{unbound}}$ is the mean unbound time calculated from simulations. (C) 3-bead model: the typical distance $\sqrt{2\epsilon T_{\textrm{unbound}}}$ beads diffuse while unbound for three different values of $\epsilon$; $T_{\textrm{unbound}}$ is the mean unbound time calculated from simulations.}
    \label{fig:timescales}
\end{figure*}

If we take the fluctuations to zero at different rates, one source of noise ``dominates" over the other in the limit as $\epsilon\to 0$. Choosing to scale $S$ by $\frac{1}{\epsilon^2}$, the fluctuations of the binding force go to zero faster than the thermal fluctuations. 
An effective binding force that allows for stable clusters remains, but the fluctuations in the binding force are now much smaller and do not interact with the thermal noise to aid in transitions between clusters.
Indeed, we recover the naive time average to describe the effective energy barrier,
which is also the limiting barrier as $\alpha\to\infty$ of the above-mentioned fluctuation-interaction case.

By keeping $S$ independent of $\epsilon$, we flip which fluctuation source dominates in the limit as $\epsilon\to 0$.  Here, the switching between Markov chain states occurs so infrequently
 that the system equilibrates while in this one state.  No effective force is generated by the superposition of different states, thus no clusters are predicted.  In the 3-bead model, two cluster-like states appear but their generating mechanism would not produce clusters in the 361-bead model.  
 The first is a 3-bead large-triangle configuration (Fig.~\ref{fig: states}A), but it is a result of the balance between the repulsive excluded volume force and confinement force. In the 361-bead model, this would correspond to all beads in an amorphic arrangement. 
 The second is a 2-bead-like cluster formed by the fact that a crosslink binds two beads; this cluster returns to the 3-bead triangle configuration once the bond is broken.  In the 361-bead model, this would correspond to another amorphic arrangement with many pair-wise bound beads.

While mathematically the three regimes are distinguished by the relative rates at which the two fluctuations go to zero asymptotically, experimentally only the binding timescale (or cluster lifetimes) and thermal diffusivity can be determined. To help bridge the gap from mathematics to the real world, we show in Fig.~\ref{fig:timescales}(A-C) the typical distance a bead diffuses during the average time the bead remains unbound for the case of (A) the biologically realistic model in \cite{hult_enrichment_2017} (B) the reduced but still dimensional 361-bead model of the nucleolus used in Fig.~\ref{fig: 361mixing} and (C) the idealized 3-bead model.  The three regimes of rigid (blue), flexible (pink), and amorphic (yellow) are added based on measurements of the clustering behavior in (A) and (B) and based on the quasipotential barrier heights in (C).  For (C), typical flexible-regime distances are approximately 0.25 to 0.35 and beads rarely form bonds when farther than 1 unit away.  Similarly, in (A), typical flexible distances are approximately 15 to 25 nm and beads cannot bind if further than 90 nm away, a similar ratio to (C).  The model in (B) appears to require faster binding to create clusters. Regardless, both models in (A) and (B) cover biologically feasible length and timescales. Note that strict boundaries cannot be drawn between regimes and these fuzzy boundaries depend on the criteria used.

The emergence of a wide range of clustering regimes (amorphic, flexible, rigid)(Fig.~\ref{fig:timescales}) and exchange within and between clusters with a pared-down 3-bead model (Fig.~\ref{fig: MPP and EB}) provides a framework for understanding the governing principles of genome organization. The crosslinking forces provide a mechanism to build gene (bead) clusters and generate informational circuits that are agnostic to the position of a gene along a given chain of chromatin.
By adjusting the timescale of the crosslinking forces, the plasticity of the gene clusters can be tuned based on biological needs. 
This is akin to tuning the crosslinking timescale between the rigid (blue) and flexible (pink) regions of Fig.~\ref{fig:timescales}.
The biological mechanism is agnostic to the division between these two regions; tuning the timescale of the crosslinking forces allows the cell to modify the rigidity of clusters.
  The arrangement of beads within clusters provides a physical model for the dynamics of gene clusters within the nucleus. The ability to spatially reconfigure genes loci is central to the cell's ability to re-wire its transcriptional circuitry. The 3-bead model shows that the energy barrier for rearrangements within a cluster is lower than that required for exchanges between clusters (Fig.~\ref{fig: MPP and EB}E, F). 
The system is remarkably plastic, with small bursts of energy able to potentiate new configurations depending on the biological demand.  
This model and analysis provide critical insight into the diversity of chromosomal geometry with a minimal set of parameters.

Computationally, our analysis is limited to relatively small numbers of beads due to the need to enumerate all possible pairwise bound states for the Markov chain.  With the growing power of machine learning, it is likely possible to build a physics-informed neural network \cite{li_machine_2022, lin_data_2022} to learn the effective potential. 
This would allow predictions of cluster dynamics for larger systems but would need to be re-applied for each timescale or set of model parameters.  The effective potential alone gives little insight into the {\em mechanism} underlying the emergent behavior that our analysis of the 3-bead model has provided.
This new outlook on the importance
of noise at the proper timescale aids in deepening our
understanding of life at the cellular
level.

\section{Methods}

The idealized chromatin model is based on the polymer-like chain of beads model in \cite{hult_enrichment_2017,walker_transient_2019}.  The position of bead $i$ is given by the overdamped Langevin equation 
\begin{equation}\label{eq: system}
    d\mathbf{x}_i = (\mathbf{f}_c^i+\mathbf{f}_{\text{EV}}^i + \mathbf{f}_{\text{bond}}^i)dt + \sqrt{2\epsilon} d\mathbf{B}.
\end{equation}
The parameter $\epsilon$ scales the vector of Brownian increments $d\mathbf{B}$ preparing for asymptotic analysis to take the fluctuations to zero.  The deterministic confinement force, 
$$
\mathbf{f}_c^i = -\eta \mathbf{x}_i,
$$
replaces the hard-wall constraint of the nucleus membrane while the excluded volume force,
$$
\mathbf{f}_{\text{EV}}^i = \sum_{j \neq i}a_{ev}(\mathbf{x}_i-\mathbf{x}_j)\exp \bigg(-\frac{|\mathbf{x}_i-\mathbf{x}_j|^2}{c_{ev}}\bigg),
$$
remains similar to its form in \cite{hult_enrichment_2017,walker_transient_2019}.  Parameters $\eta$, $a_{ev}$, and $c_{ev}$ are given in Table 1 for both the 361-bead $(\mathbf{x}_i\in\mathbb{R}^3, i=1\dots 361)$ and the 3-bead $(\mathbf{x}_i\in\mathbb{R}^2, i=1\dots 3)$ models.  Note we have neglected the spring force linking the beads together to form a chain. We show in Fig.~\ref{fig: 361mixing} that this spring force is not necessary to produce clustering dynamics.  

The stochastic binding force,
$$
\mathbf{f}_{\text{bond}}^i = \sum_{j \neq i} \kappa b_{ij}(\mathbf{x}_j-\mathbf{x}_i),
$$
models the binding SMC proteins found in the biological system. This force binds two beads, corresponding to $b_{ij} = 1$ if beads $i$ and $j$ are bound and 0 otherwise.
Each bead can be bound to only one other bead; these stochastic bonds form and break at exponentially distributed times with a binding rate proportional to the bead separation distance, $a(\mathbf{x}_i-\mathbf{x}_j)$, and constant breaking-rate $c$. 
The time evolution of the states follows a general CTMC process; for the 3-bead model the transition rate matrix is given by 
\begin{equation}\label{CTMC}    
 S =  \begin{pmatrix}
    b & c & c & c \\
    a(\mathbf{x}_1-\mathbf{x}_2) &-c& 0 &0 \\
    a(\mathbf{x}_1-\mathbf{x}_3) & 0 & -c &0 \\
    a(\mathbf{x}_2-\mathbf{x}_3) & 0 & 0 & -c
    \end{pmatrix}
\end{equation}
with $b = -a(\mathbf{x}_1-\mathbf{x}_2)-a(\mathbf{x}_1-\mathbf{x}_3)-a(\mathbf{x}_2-\mathbf{x}_3)$. 

The affinity function $a(\mathbf{x})$ and parameter values are given in Table 1 for both the 361-bead and the 3-bead models.  Both $a(\mathbf{x})$ and $c$ are scaled by $\beta$ for the 361-bead model and $\alpha$ for the 3-bead model to explore the kinetic timescale on which the crosslinks bind and unbind, mirroring the kinetic timescale parameter $\mu$ in the large-scale simulations of \cite{walker_transient_2019}.  Both $a(\mathbf{x})$ and $c$ are further scaled by different powers of $\epsilon$ to perform the asymptotic analysis of the fluctuations.

\begin{table}[t!]
\centering
\caption{Model parameters and formula.}
\begin{tabular}{lrr}
 & 361-Bead Model & 3-Bead Model  \\
\midrule
$ a(\mathbf{x})$ & $ \displaystyle\frac{2}{1+\exp({20 ( |\mathbf{x}|-75)})}$ &    $\displaystyle \frac{2}{1+\exp({20 ( |\mathbf{x}|-0.75))}}$ \\ 
 $\eta$ & 0.002 \; $s^{-1}$ & 1 \\
$a_{ev}$ & 0.03 \; $s^{-1}$ & 2  \\
$c_{ev}$ & 30,000 \; $nm^2$ & 0.5 \\
$\kappa$ &10 \; $s^{-1}$ & 5 \\
$c$ & 0.01 \; $s^{-1}$ & 0.5  \\
\bottomrule
\end{tabular}
\end{table}

\subsection*{Monte Carlo Simulations}

We perform Monte Carlo simulations of the 3-bead system to compare the scaling of the escape times with $\epsilon$ to the effective energy barriers predicted by the theoretical calculations.  The simulations are started in either the 2-bead or 3-bead cluster, and continued until a stopping condition is met, indicating that the system has left the basin of attraction of the initial state.  From the 2-bead cluster, we look for either a different 2-bead cluster or the 3-bead triangle cluster.  From the 3-bead triangle cluster, we look for either a 2-bead cluster or a collinear configuration that is the saddle point between different arrangements of the 3-bead triangle cluster.
The mean escape time, $\tau$, is computed by the maximum likelihood estimate that divides the sum of all escape times by the number executing the desired transition.

Code for reproducing results is available
on GitHub 
\cite{bwalker1_doi_2022}
(\verb^quasi-string-reprod^).

\section*{Acknowledgements}

AC and KN were partially supported by National Science Foundation (NSF) grants DMS-1816394 and DMS-2307297 and AC was also partially supported by NSF grant DMS-1929298.  KB was partially supported by the National Institutes of Health (NIH) grant R01 GM32238.


%

\end{document}